\documentclass{article}

\usepackage{times}
\usepackage{graphicx} 

\usepackage{natbib}

\usepackage[plain]{algorithm}
\usepackage{algorithmic}

\usepackage{amsmath}
\usepackage{amssymb}

\usepackage{hyperref}

\usepackage[accepted]{icml2017}

\usepackage{enumitem}
\usepackage[framemethod=tikz,xcolor=true]{mdframed}
\usetikzlibrary{shadows}
\newmdenv[%
    nobreak=true,leftmargin=0.2cm,%
    backgroundcolor=yellow!10,%
    roundcorner=5pt,%
    frametitlefont=\normalfont,%
    frametitlerule=true,%
    font=\fontsize{8}{10},%
    tikzsetting={draw=black, line width=1.0pt},%
    shadow=true,shadowsize=6pt%
    ]{Algoritmo}%

\icmltitlerunning{Active Preference Learning for Personalized Portfolio Construction}

\usepackage{amsmath}
\usepackage{color}
\usepackage{wrapfig}
\DeclareMathOperator*{\argmax}{arg\,max}

\newcommand{\uu}{{\mathbf{u}}}

\newcommand{\ww}{{\mathbf{w}}}
\newcommand{\xx}{{\mathbf{x}}}

\newcommand{\xxopt}{{\mathbf{x}_{\text{opt}}}}
\newcommand{\yy}{{\mathbf{y}}}
\newcommand{\zz}{{\mathbf{z}}}

\newcommand{\RR}{\mathbb{R}}

\newcommand{\algref}[1]{\hyperref[#1]{Algorithm~\ref{#1}}}
\newcommand{\figref}[1]{\hyperref[#1]{Figure~\ref{#1}}}
\newcommand{\tabref}[1]{\hyperref[#1]{Table~\ref{#1}}}
\newcommand{\secref}[1]{\hyperref[#1]{Section~\ref{#1}}}

\begin{document} 
	
\twocolumn[
\icmltitle{Active Preference Learning for Personalized Portfolio Construction}



\icmlsetsymbol{equal}{*}

\begin{icmlauthorlist}
	\icmlauthor{Kevin Tee}{sigopt}
	\icmlauthor{Michael McCourt}{sigopt}
	\icmlauthor{Ruben Martinez-Cantin}{sigopt,zaragoza}
	\icmlauthor{Ian Dewancker}{uber}
	\icmlauthor{Frank Liu}{waterloo}
\end{icmlauthorlist}

\icmlaffiliation{sigopt}{SigOpt, San Francisco, California, USA}
\icmlaffiliation{uber}{Uber Technologies, San Francisco, California, USA}
\icmlaffiliation{waterloo}{University of Waterloo, Waterloo, Ontario, Canada}
\icmlaffiliation{zaragoza}{Centro Universitario de la Defensa, Zaragoza, Spain}

\icmlcorrespondingauthor{Kevin Tee}{kevin@sigopt.com}

\icmlkeywords{Bayesian optimization}

\vskip 0.3in
]



\printAffiliationsAndNotice{}  

\begin{abstract}
	In financial asset management, choosing a portfolio requires balancing returns, risk,
	exposure, liquidity, volatility and other factors.  These concerns are difficult
	to compare explicitly, with many asset managers using an intuitive or implicit sense
	of their interaction.  We propose a mechanism for learning someone's sense of
	distinctness between portfolios with the goal of being able to identify portfolios which are
	predicted to perform well but are \emph{distinct} from the perspective of the user.
	This identification occurs, e.g., in the context of Bayesian optimization of a backtested
	performance metric.  Numerical experiments are presented which show
	the impact of personal beliefs in informing the	development of a diverse and high-performing portfolio.
\end{abstract}

\section{Introduction\label{sec:introduction}}

Many problems of optimal decision fall in the context of nonlinear
optimization, where the goal is to find the maximum of a function $f:\Omega\to\RR$ over some domain $\Omega\subset\RR^s$,
\begin{align}
	\label{eq:optimizationproblem}
	\xxopt = \argmax_{\xx\in\Omega} f(\xx).
\end{align}
In this context, the optimal decision results in a single value
$\xxopt$. However, many applications such as financial management
may benefit from finding multiple $\xx^*$ locations with high $f(\xx^*)$ values that are
also sufficiently distinct from each other (in some sense)
so as to build a diversified portfolio. Finding multiple optimal decisions
which are suitably distinct presents new challenges. First, we must
define the \emph{distinctness} between selections of $\xx$. In this
work, the ability to sense relative distinctness between proposed portfolios is learned directly from the user
by querying the financial expert on the distinctness of a proposed $\xx$
as compared to $\xxopt$, which itself can be found through standard black-box
optimization.

Thus, the optimal decision problem can be considered as the problem of identifying a
\emph{supplemental set} of financial portfolios $\{\xx_{e1},\ldots,\xx_{em}\}$ which balance a high
value against a degree of distinctness from $\xxopt$. The user may choose from among these to supplement
$\xxopt$ in the eventual portfolio (thus these need not be local optima, nor
within any specified range).

Finding multiple solutions to an optimization problem has been
previously studied. The idea of multimodality has been extensively explored
in the evolutionary algorithms literature; see, e.g.,
\cite{DBLP:journals/corr/Wong15} and references therein. In the
context of Bayesian optimization, other authors have explored this
idea by finding multiple local optima for robustness
\cite{GuentherLeeEtAl14, NogueiraIROS16} or parallelization
\cite{SnoekLarochelleEtAl12, GinsbourgerEtAl07, NguyenNIPS16}
Another interesting area of finding diverse sets of solutions is the determinantal point
process \cite{KuleszaTaskar12}.


Along the path of recent advances in interactive learning in the context of
optimization, this work relies on Bayesian optimization for nonlinear
optimization because it is sample efficient.
For decision making and preference learning, this
sample efficiency is directly translated in a resource-efficient
optimization in terms of computation and user queries. Following the
same paradigm, \cite{DewanckerEtAl16} proposed to learn the user's
preference among a set of utility functions in multimetric
optimization, \cite{BrochuEtAl08, BrochuEtAl10} used this
concept for learning virtual materials and smoke simulations,
\cite{ThatteICRA17} applied the same approach for prothesis design and
\cite{Okuma2011} used it for image classification. 
Recently,
\cite{GonzalezNIPS16}
presented an alternative to preference learning
based on Copeland functions.

\section{Using Preferences to Infer Distinctness\label{sec:distancedef}}

Our goal is to build a model which, if given portfolios $\ww,\xx,\yy,\zz\in\Omega$, can
estimate
\begin{align}
	\label{eq:probabilitydef}
	\Pr(d(\ww, \xx) > d(\yy, \zz)),
\end{align}
where $d:\Omega\times\Omega\to\RR_+$ is the user's implicit sense of distinctness.  For
practical purposes, $d$ can be thought of as a distance, although $d$ is not required to
satisfy the \emph{triangle inequality}.

Therefore, using the model does not give an explicit distance between
two portfolios; instead, it returns the probability that two portfolios
are more distinct than two other portfolios. This allows for a
flexible sense of distinctness as an explicit form need not be
specified \textit{a priori}.

This model is equivalent to a binary classification problem. Thus, we
use logistic regression to model this probability. The input features are
$\ww,\xx,\yy,\zz\to\begin{pmatrix}|\ww-\xx|\\|\yy-\zz|\end{pmatrix}$,
the concatenation of $|\ww-\xx|$ and $|\yy-\zz|$,
where $|\uu|$ is the elementwise absolute value of a vector $\uu$.

\begin{algorithm}[ht]
  \centering
  \begin{Algoritmo}[frametitle={Input: $n$ number of iterations, $m$
      number of elements to rank, $\alpha$}]
    {\scriptsize
        \begin{itemize}[noitemsep,nolistsep]
  \item Run standard Bayesian optimization for $n$ iterations to find
    $\xxopt$
    \begin{itemize}[noitemsep,nolistsep]
    \item At each iteration, observe the point with highest EI
    \end{itemize}
  \item Initialize classifier
  \item Run parallel Bayesian optimization for $n$ iterations
  \begin{enumerate}[noitemsep,nolistsep]
  \item At each step, select $m$ points $\{\xx_1 \ldots \xx_m\}$with
    highest EI
  \item \label{alg:rank}  Query the user to rank $d(\xxopt, \xx_i)$
	  for $1\leq i\leq m$
    \item Observe $f(\xx_i)$ for $\xx_i$ ranked most distinct in step~\ref{alg:rank}
        \item Update classifier from rankings and $\xxopt$, if needed.
    \end{enumerate}
  \item Return the $\alpha$-\emph{distinctly efficient portfolio} using the learned classifier
  \end{itemize}}
  \end{Algoritmo}
  \caption{Adaptive preference learning within Bayesian optimization for portfolio construction.}
  \label{fig:algorithm}
\end{algorithm}

\section{Active Preference Sampling within Bayesian Optimization\label{sec:guidedoptimization}}

\algref{fig:algorithm} shows the proposed algorithm. As discussed
earlier, $\xxopt$ is determined absent any input from the user: it is
simply the global maximum of the portfolio performance function $f$.
To find the supplemental distinct ``solutions'' to the optimization
problem we leverage the user's sense of distinctness
to inform decisions made within a Bayesian optimization loop.

First, we perform a standard Bayesian optimization loop where, at each
iteration, a new suggestion $\xx_s$ is selected by maximizing the
expected improvement (EI) function
\cite{JonesSchonlauEtAl98}.  After the initial optimization for $\xxopt$,
we focus on the problem of learning/exploiting distinctness.  In this
case, we continue with the Bayesian optimization, but generate $m$
solutions in parallel using the constant liar mechanism
\cite{GinsbourgerEtAl07}. During this supplemental search, the next suggestion $\xx_s$ is
defined as the portfolio chosen to be most distinct by the user.

At each iteration,
the rankings are converted into $2\binom{m}{2}$ pairwise
classification data points for the logistic model.  For
example, if the customer responds that $\xx_1$ should be ranked as
less distinct from $\xxopt$ than $\xx_2$, we now have the data
\begin{align*}
	\xxopt,\xx_1,\xxopt,\xx_2,&\;\;\text{False}, \\
	\xxopt,\xx_2,\xxopt,\xx_1,&\;\;\text{True}.
\end{align*}

The choice of the number of ranked portfolios $m$ impacts the
search for distinctness: fewer portfolios will bias towards
exploitation and more portfolios will yield more exploration.
More analysis on this impact will benefit the efficiency of
the search process. For the experiments in \secref{sec:experiments}
we choose $m=5$ which, while chosen somewhat arbitrarily, proves to be a reasonable
balance between the amount of data provided from a query and
difficulty for a user to resolve that query.

Although we have defined a sequential algorithm where the user is
queried after each iteration of the Bayesian optimization algorithm,
if the user has accessibility constraints, the user may respond to
batch queries immediately after determining $\xxopt$ to fully prepare
the model for the distinctness search.  This is the "Initialize classifier" step,
and can be guided by principles similar to \cite{ThatteICRA17} or
\cite{DewanckerEtAl16}.

\section{Selecting Portfolios after the Search\label{sec:wholealgorithm}}

Thus far, our process consists of identifying the globally optimal portfolio $\xxopt$,
and then searching for supplemental high-performing portfolios $\xx_{e_1},\ldots,\xx_{e_m}$
which enjoy varying levels of distinctness from $\xxopt$ based on the user's preferences
(as learned through queries administered either before or during the supplemental search).
To resolve a trading strategy based on these supplemental portfolios, we try to address
the multicriteria optimization problem
\begin{align}
	\label{eq:mcproblemdef}
	\begin{split}
		\max_{\xx\in\Omega} &\quad f(\xx), \\
		\max_{\xx\in\Omega} &\quad d(\xx, \xxopt).
	\end{split}
\end{align}

Without the actual $d$ (we know only the preference model for \eqref{eq:probabilitydef}),
we can only resolve our efficient set of points (for which $f$ can not be improved without
decreasing $d$) in some partial sense.
Let us denote all the suggested portfolios during the supplemental search as
$\xx_1,\ldots,\xx_n$ and state, without loss of generality, that
\[
	f(\xx_1)\geq\cdots\geq f(\xx_n).
\]
We define $\xx_i$ to be an $\alpha$-\emph{distinctly efficient portfolio} if
\begin{align*}
	\Pr(d(\xxopt, \xx_i) > d(\xxopt, \xx_j)) > \alpha,\quad 1< j < i.
\end{align*}
As the base case, we always include $\xx_1$ in the efficient set.

Because portfolios are indexed by nonincreasing function value, a portfolio
is $\alpha$-distinctly efficient
only if the model predicts its distinctness from $x_{opt}$ is greater than each
point already in the efficient set that preceded it with probability at least $\alpha$.
This $\alpha$ value is a free parameter for the user to choose when deciding which
portfolios to implement: higher values exclude portfolios which are less certainly distinct.

\section{Numerical Experiments\label{sec:experiments}}

The data for our experiments is pulled from the S\&P 500 during 2016. Our trading strategy
can take long positions in 5 aggregated industries
(industrials, energy, consumer discretionary, utilities, telecommunications).
For a specific day on which a trading strategy is desired, the $f$ defining
\eqref{eq:optimizationproblem} is the Sharpe ratio of a given portfolio for the
previous 10 trading days, estimated with discrete 1 day intervals.

A portfolio is a partition of unity (i.e., $\xx>0$ and $\|\xx\|_1=1$), which must be enforced
during the Bayesian optimization.  To do this, we solve an adjacent problem involving
$\tilde{\xx}\in\RR_+^4$ such that our portfolio is
\[
	\xx = \frac{1}{1 + \|\tilde{\xx}\|_1}\begin{pmatrix}\tilde{x}_1\\\tilde{x}_2\\\tilde{x}_3\\\tilde{x}_4\\1\end{pmatrix}.
\]
The optimization problem is limited to $\Omega=[10^{-3},10^3]^4$ and is solved on a log scale.
This allows us to approximately search the space of partitions of unity for $\xx\in\RR_+^5$.

We run the optimization loop for a total of 60 iterations to determine $\xxopt$.
Then a second 60 iteration length
Bayesian optimization augmented with the user's preferences as described in \secref{sec:guidedoptimization}
to determine a set of $\xx_1,\ldots,\xx_{60}$ points from which we can identify a
set of $\alpha$-distinctly efficient portfolios.

\subsection{Visualizing the Efficient Points\label{sec:changingalpha}}

\begin{figure}[ht]
	\centering
	\includegraphics[width=\columnwidth]{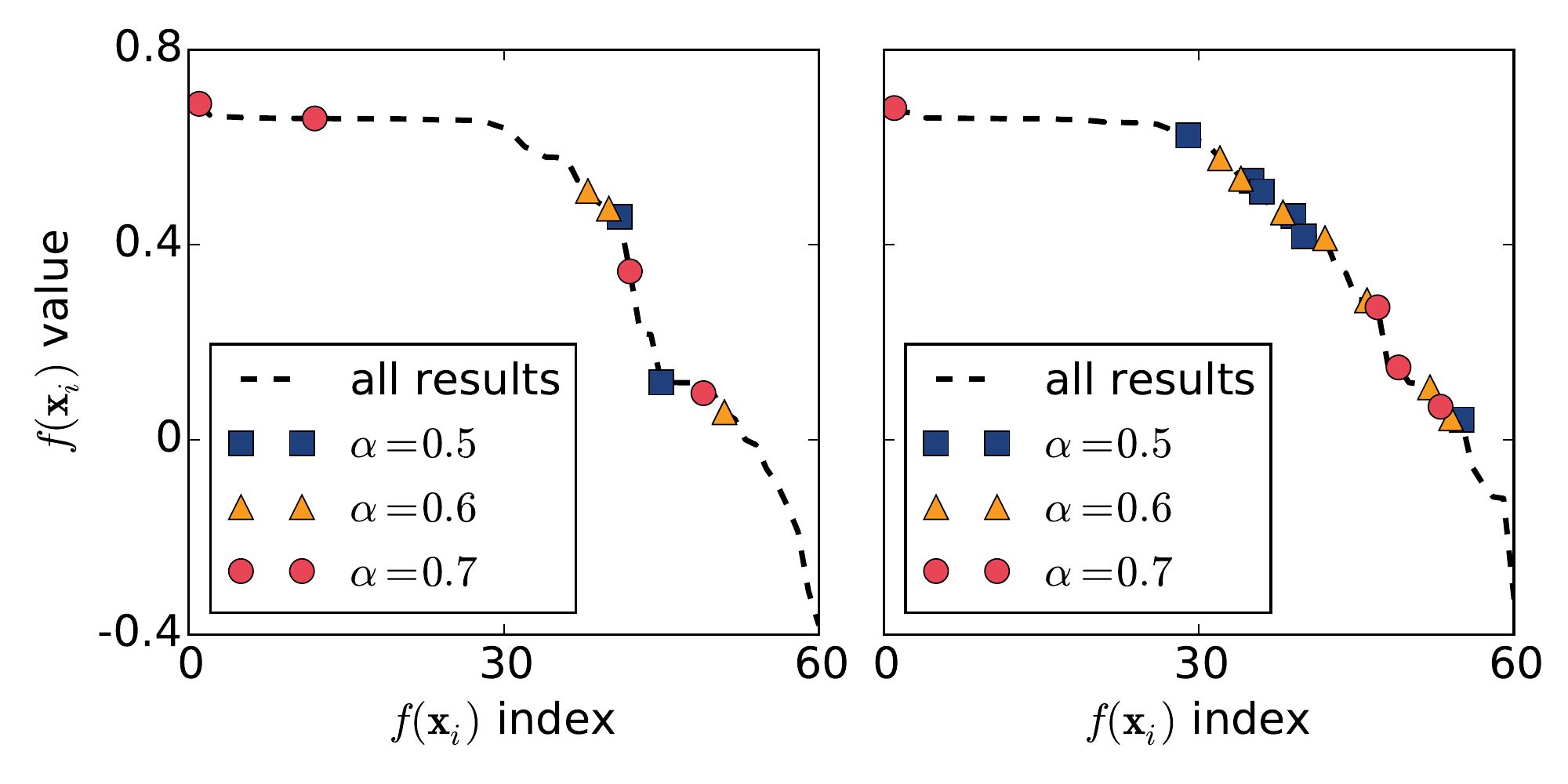}
	\caption{As $\alpha$ decreases, more of the results are included,
		e.g., portfolios distinct for $\alpha=.7$ are a subset of those distinct for $\alpha=.6$.
		\textit{left}: A sense of distinctness proportional to the Euclidean distance.
		\textit{right}: Distinctness among utilities and telecommunications assets is preferred.
		\label{fig:efficientsetcomparison}}
\end{figure}

Visualizing the solutions to \eqref{eq:mcproblemdef} would normally be managed
graphically with the Pareto frontier.  In this setting, however, a standard frontier is unavailable
because we have only the ability to estimate \eqref{eq:probabilitydef} and not to compute $d$.
Furthermore, different choices of $\alpha$ will change the portfolios which appear in the
efficient set.  \figref{fig:efficientsetcomparison} shows the impact of various $\alpha$ values
when identifying $\alpha$-distinctly efficient portfolios with different preferences.

The first point to recognize in \figref{fig:efficientsetcomparison} is that, because different
preferences search the domain differently, the distribution of portfolio performances $f(\xx)$
will vary; this is why the dashed line representing all 60 observed function values varies.
The second point to note is that the portfolios distinct for a given $\alpha$ value include
all the portfolios distinct for any greater $\alpha$ values.  As such, the user has one final
decision, albeit slightly outside the loop, in choosing the $\alpha$ value to identify
which of the portfolios from all the results are to be used in effecting a trading strategy.

\subsection{Evaluating Portfolio Performance\label{sec:differentstrategies}}

In this section we study the impact of using $\alpha$-distinctly efficient portfolios
suggested by the strategy above to execute trades on a specific day.  We build portfolios
on the second Wednesday of each month of 2016, as formed based on the previous 10
trading days; January was excluded so as to use only data from 2016.
The trading strategy consists of $\xxopt$ supplemented
equally by the $m$ distinct portfolios,
\[
\frac{1}{2}\left(\xxopt + \frac{1}{m}(\xx_{e_1} + \ldots + \xx_{e_m})\right).
\]
Recall that $\xx_{e_1}, \ldots, \xx_{e_m}$ are impacted by the choice of $\alpha$.

\begin{figure}[ht]
	\centering
	\includegraphics[width=\columnwidth]{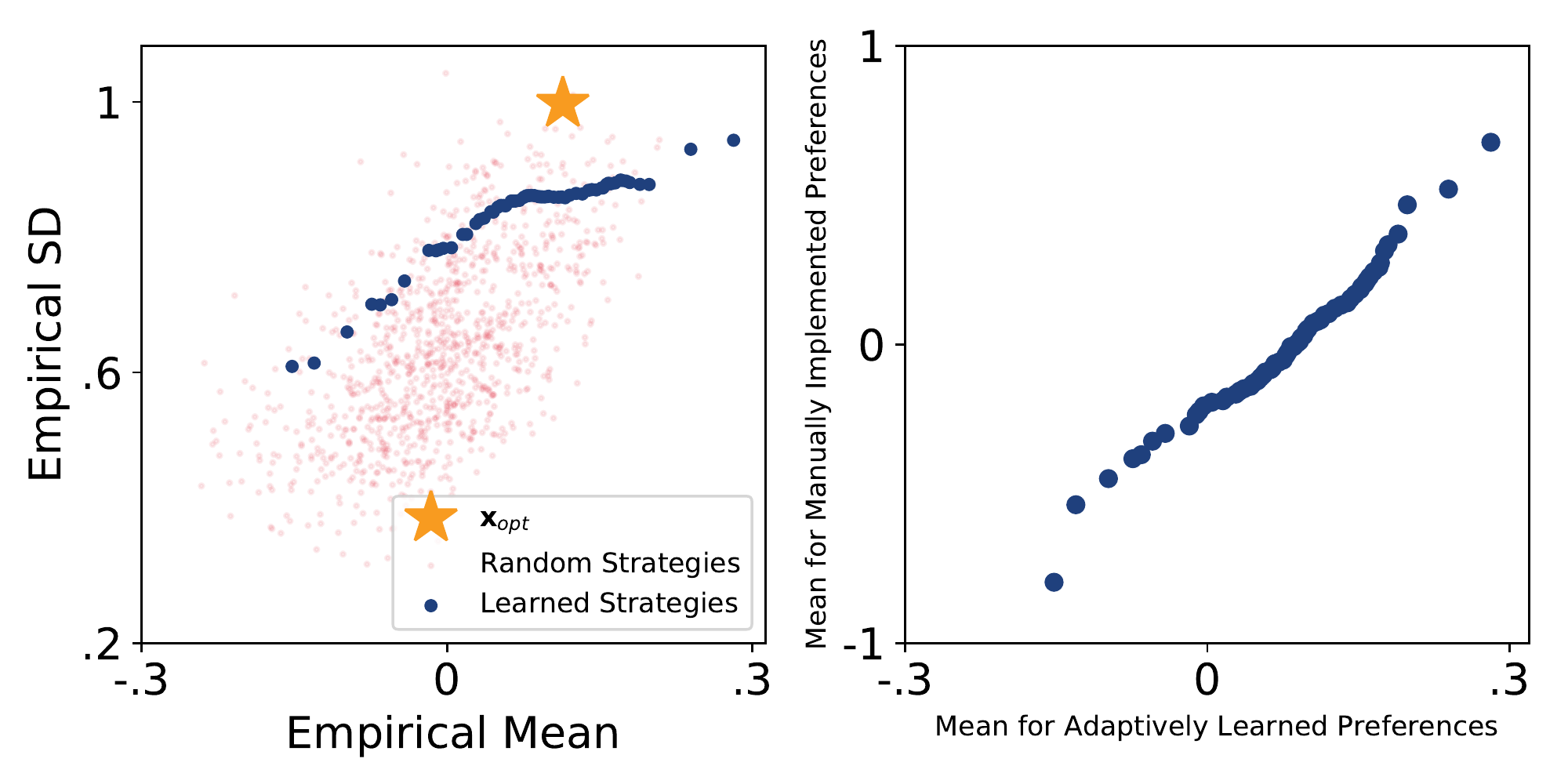}
	\caption{\textit{left}: We can study the statistics of trading on the 11 trading days mentioned
		and notice that by supplementing $\xxopt$ with additional distinct portfolios as learned
		during the optimization we have the potential to decrease the variance without negatively
		impacting the mean.
		\textit{right}: There is a strong positive correlation between trading strategies manually
		filled with high-performing assets and those trained based on a preference for
		distinctness among high-performing assets, regardless of the proportion of high-performing assets.
		\label{fig:meanvariance}}
\end{figure}

The performance of a trading algorithm is judged by both the empirical mean and variance
observed over the 11 trading days.
The left half of \figref{fig:meanvariance} shows the performance of a trading strategy based
solely on $\xxopt$ as well as possible trading strategies learned for various preferences
regarding distinctness.  In particular, supplementing the $\xxopt$ portfolio with alternate
portfolios generally has the effect of reducing variance, but it has the potential to do so without
negatively impacting the mean (which would be the case if supplemental portfolios were added
randomly as seen in the spread of random outcomes).

The right half of \figref{fig:meanvariance} shows the conditions under which variance can be
reduced without impacting the mean, namely whether the user prefers distinctness among assets that
will perform well.
This graph shows a strong positive correlation in empirical mean for portfolios
manually loaded with high performing assets and those which are only built
with a preference for distinctness in high performing assets.

The important factor here is that we take \emph{no explicit action} to cause the eventual trading strategy
to contain high performing assets; we only encourage exploration of the performance function $f$
on the domain $\Omega$ with a desire for distinctness as preferred by the user.
Essentially, if the user is able to successfully identify assets which will perform well, $\xxopt$ can
be supplemented in such a way as to reward this belief by decreasing the variance without
decreasing the mean.

\section{Future Work\label{sec:futurework}}
The impact of varying $\alpha$ in identifying the final trading strategy is unclear, especially
given that, with enough information, the user's preferences in all possible pairwise
comparisons can be learned.  Future work must determine how best to transition the
results of the supplemental search into a well-balanced trading strategy.  Furthermore, we should
analyze the necessary balance between working on identifying $\xxopt$ and the supplemental
portfolios as evaluating a proposed portfolio could be both costly and time consuming.
Additionally, the impact of batch querying after identifying $\xxopt$ has not yet been studied,
but would likely be significant in a practical implementation.



\bibliographystyle{icml2017}

\end{document}